\documentclass[useAMS,usenatbib]{mn2e}

\usepackage{aas_macros}
\usepackage{graphicx}
\usepackage{ulem}
\usepackage{lscape}
\usepackage{mathptmx}

\def\Lya{\mbox{Ly$\alpha$}}
\def\lymana{\mbox Lyman $\alpha$}

\newcommand{\lya}{Ly$\alpha$}

\newcommand{\kms}{\,\,{\rm km / s}}

\def\HI{\mbox{H\,{\sc i}}}

\def\CIV{\mbox{C\,{\sc iv}}}
\def\CIVa{\mbox{C\,{\sc iv}a}}
\def\CIVb{\mbox{C\,{\sc iv}b}}

\def\MgII{\mbox{Mg\,{\sc ii}}}

\def\AlII{\mbox{Al\,{\sc ii}}}
\def\AlIII{\mbox{Al\,{\sc iii}}}

\def\SiIV{\mbox{Si\,{\sc iv}}}

\def\SiIVb{\mbox{Si\,{\sc iv}b}}
\def\FeII{\mbox{Fe\,{\sc ii}}}

\title[The \CIV\ forest as a probe of BAO]{The \CIV\ Forest as a probe of baryon acoustic oscillations}
\author[Pieri M. M. et al.]
{Matthew M. Pieri\thanks{E-mail:matthew.pieri@lam.fr}
 \\
 Institute of Cosmology \& Gravitation, University of Portsmouth, Dennis Sciama Building, Portsmouth PO1 3FX, UK\\
 Aix Marseille Universit\'e, CNRS, LAM (Laboratoire d'Astrophysique de Marseille) UMR 7326, 13388, Marseille, France
}

\begin{document}

\date{Accepted xxxx}

\pagerange{\pageref{firstpage}--\pageref{lastpage}} \pubyear{x}

\maketitle

\label{firstpage}

\begin{abstract}

In light of recent successes in measuring baryon acoustic oscillations (BAO) in quasar absorption using the \lymana\ (\lya) transition, I explore the possibility of using the 1548\AA\ transition of triply ionized carbon (\CIV) as a tracer. While the \Lya\ forest is a more sensitive tracer of intergalactic gas, it is limited by the fact that it can only be measured in the optical window at redshifts  $z > 2$. Quasars are challenging to identify and observe at these high redshifts, but the \CIV\ forest can be probed down to redshifts $z\approx1.3$, taking full advantage of the peak in the redshift distribution of quasars that can be targeted with high efficiency.

I explore the strength of the \CIV\ absorption signal and show that the absorbing population on the red side of the \Lya\ emission line is dominated by \CIV\ (and so will dominate over the potential BAO signal of other metals). As a consequence, I argue that forthcoming surveys may have a sufficient increase in quasar number density to offset the lower sensitivity of the \CIV\ forest and provide competitive precision using both the \CIV\ autocorrelation and the \CIV-quasar cross correlation at $\langle z \rangle \approx 1.6$.

\end{abstract}


\section{Introduction}

Baryon acoustic oscillations (BAO) are imprinted on large-scale structures and provide a probe of cosmic expansion through their use as a standard ruler (\citealt{2013PhR...530...87W} and references therein). Since the intergalactic medium is thought to represent the overwhelming majority of baryons at $z>1$ (\citealt{2009RvMP...81.1405M} and references therein), absorbers tracing these baryons are a probe of the characteristic BAO scale.
Recent results have demonstrated this by using the autocorrelation of \lya\ forest absorbers along the line of sight to distant quasars \citep{2013A&A...552A..96B,2013JCAP...04..026S, 2014arXiv1404.1801D}. They measure the BAO scale with a uncertainty of 2\% at $\langle z \rangle=2.3$ and are particularly effective at  constraining the Hubble parameter. This has been supplemented by the cross-correlation between \Lya\ forest absorption and quasars at the same redshift \citep{2014JCAP...05..027F}, which is more effective as a probe of the angular diameter distance. Together they have proven to be a powerful cosmological probe \citep{2014arXiv1404.1801D}.

The BAO peak has been observed at $z \le 0.6$ using the galaxy-galaxy correlation function (e..g. \citealt{2011MNRAS.415.2892B,2014MNRAS.439...83A}), but a gap exists at intermediate redshifts. This will be filled by the extended Baryon Oscillation Spectroscopic Survey (eBOSS) through surveys of luminous red galaxies, emission line galaxies and quasars.
An additional, and effectively free probe, is possible at $1.3<z<3.5$. \Lya\ forest based studies are not possible at $z<2$ as \Lya\ leaves the optical band, but metal forests with longer transition wavelengths allow the possibility of tracing structure at lower-redshifts. Triply ionized carbon (\CIV) provides a useful doublet \CIVa\ ($\lambda 1548$\AA) and \CIVb\ ($\lambda 1551$\AA) as it is an effective tracer of metal enriched gas over a wide range of densities (e.g. {\citealt{2003ApJ...596..768S}), has high oscillator strength, and a suitable rest-frame wavelength. 

In this Letter, I shall explore the potential for using the \CIV\ forest to measure the BAO scale and so provide a new probe of the expansion of the universe. This work is set out as follows: in Section~\ref{data}, the data set used will be described, in Section~\ref{strength}, the strength of the \CIV\ signal is explored, Section~\ref{metalpop} describes tests of the metal population, Section~\ref{survey} presents a potential survey, and Section~\ref{challenges} explores some observational challenges.

\section{Data}
\label{data}

I use the Sloan Digital Sky Survey III Data Release 9 \citep{2012ApJS..203...21A} quasar sample
compiled by visual inspection of candidates as outlined in \cite{2012A&A...548A..66P}. 
 \Lya\ absorption with a  redshift range $2<z<3.5$ and a band-wise signal-to-noise ratio, S/N$>8$ (see \citealt{2014MNRAS.441.1718P}, P14 hereafter) is taken from the high redshift sample of 54 468 quasars (as detailed in \citealt{2013AJ....145...69L}) to explore relative strengths of \Lya\ and \CIV\ forests as described in Section~\ref{strength}. A broader  sample of quasars, including 9523 at low redshifts taken from Data Release 9, was used to blindly explore absorption on the red side of the \Lya\ emission line corresponding to \CIV\  with redshifts $1.3<z< 3.5$ (also limited to a band-wise S/N$>8$). These spectra were prepared using the spline continuum fitting procedure outlined in P14.

In exploring the red side bands, I define two bands in the quasars' rest-frame, 1260--1375\AA\ (with both \CIV\ and \SiIV\ absorption; the `\CIV+\SiIV\ band') and 1400--1520\AA\ (without \SiIV\ absorption;  the `\CIV\ band'). The boundaries of these bands are conservatively defined to avoid uncertainly in fitting \Lya, \SiIV\ and \CIV\ emission lines.

\section{The strength of metal absorption signal}
\label{strength}

I use two methods for characterizing the comparable signal strength between \Lya\ and the \CIV\ forests. The most direct statistical comparison of the bulk metal signal and the bulk \Lya\ forest is presented by measurements of the 1D power in both the \Lya\ forest and on the red side of the \Lya\ emission line. Fig. 19 of \citet{2013A&A...559A..85P} shows the power in the two restframe bands of interest, while fig. 20 shows the 1D power spectrum measured in the forest at various redshifts. In comparing these two figures one can see that the metal power in these red side bands is a factor of 5--20 lower than that of the forest power on scales greater than that of the doublet  ($k < 0.006 (\kms)^{-1}$) at 4000\AA, the mean wavelength of current BAO measurements from \Lya. This corresponds to  $z=2.3$ in the \Lya\ forest or $z=1.6$ in the \CIV\ forest.  It is also notable when comparing the power spectrum of each red side band that while there is more power in the \CIV+\SiIV\ band than \CIV\ band, the difference indicates that the additional \SiIV\ power is no greater than 50\% of that present in the \CIV\ band. 

We can also take a more focused approach by comparing the primary measured quantity in \Lya\ forest BAO studies, the flux contrast,
$\delta(\lambda)=f (\lambda) / (C (\lambda) {\bar{F}(z)}) -1$,
for both the \CIV\ and the \Lya\ forests, where $f(\lambda)$ and $C (\lambda)$ are the quasar flux and the absorbed continuum level of the quasar respectively in the relevant wavelengths ranges,  $F=f / C$ is the transmitted flux fraction (referred to as `flux' hereafter), and $\bar{F}(z)$ is the mean flux in the appropriate forest. This requires the \CIV\ to be present at the same redshift as the \Lya\ absorption and so is a test of \CIV\ at  $z>2$.

This is done by stepping through the \Lya\ forest and determining the flux contrast for \CIV\ at the same redshift. The flux contrast of this pixel pair is then calculated by renormalizing locally. Analogous to the locally calibrated pixel search of \citet{2010ApJ...716.1084P} and the composite spectra approach (\citealt{2010ApJ...724L..69P}; P14), this brute force method cancels the contribution of contaminating absorption in the local renormalization. The average $\delta_{\rm CIV}$ is calculated in bins of $\delta_{\rm Ly\alpha}$.
Fig. ~\ref{contrast} compares these quantities for the full range of $\delta_{\rm Ly\alpha}$. This indicates that the \CIV\ forest signal is between 3 and 20\% of the \Lya\ forest signal over most of the dynamic range in $\delta_{\rm Ly\alpha}$. Where the \Lya\ signal is strongest, the proportionate strength of the \CIV\ signal is also strongest, corresponding to higher density and potentially higher bias systems. This provides a conservative assessment of the \CIV\ forest signal strength as it neglects scatter in absorption strength and implicitly assumes that any inhomogeneity in \CIV\ absorption strength is stochastic. We shall return to this point in Section~\ref{survey}.

\begin{figure}
\begin{center}
\includegraphics[angle=0, width=.5\textwidth]{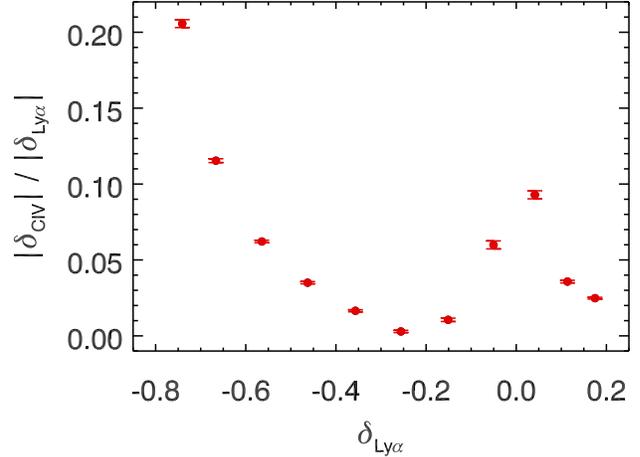}
\end{center}
\caption{The ratio of the flux contrast magnitudes for the \CIV\ and \Lya\ forests. The \CIV\ forest is typically 3--20\% weaker than the \Lya\ forest.  The stationary points in this ratio occur as absorbers with mean transmission in the \Lya\ forest have excess transmission in the \CIV\ forest. The relative \CIV\ signal is strongest for strong \Lya\ absorbers.}
\label{contrast}
\end{figure}

I conservatively conclude that the signal power in the \CIVa\ forest is a factor of 5--30 lower than that of the \Lya\ forest at $2<z<3.5$ from direct observations of these two transitions. This is consistent with the measured total power  in red side bands of \citet{2013A&A...559A..85P} at $z=1.6$, implying that the \CIVa\ forest dominates these bands. In the following section I test for other metal absorbers present in the red side bands in order to test this inference, and explore their potential for adding systematic errors.

\section{metal populations}
\label{metalpop}

\begin{figure*}
\begin{center}
\includegraphics[angle=0, width=.46\textwidth]{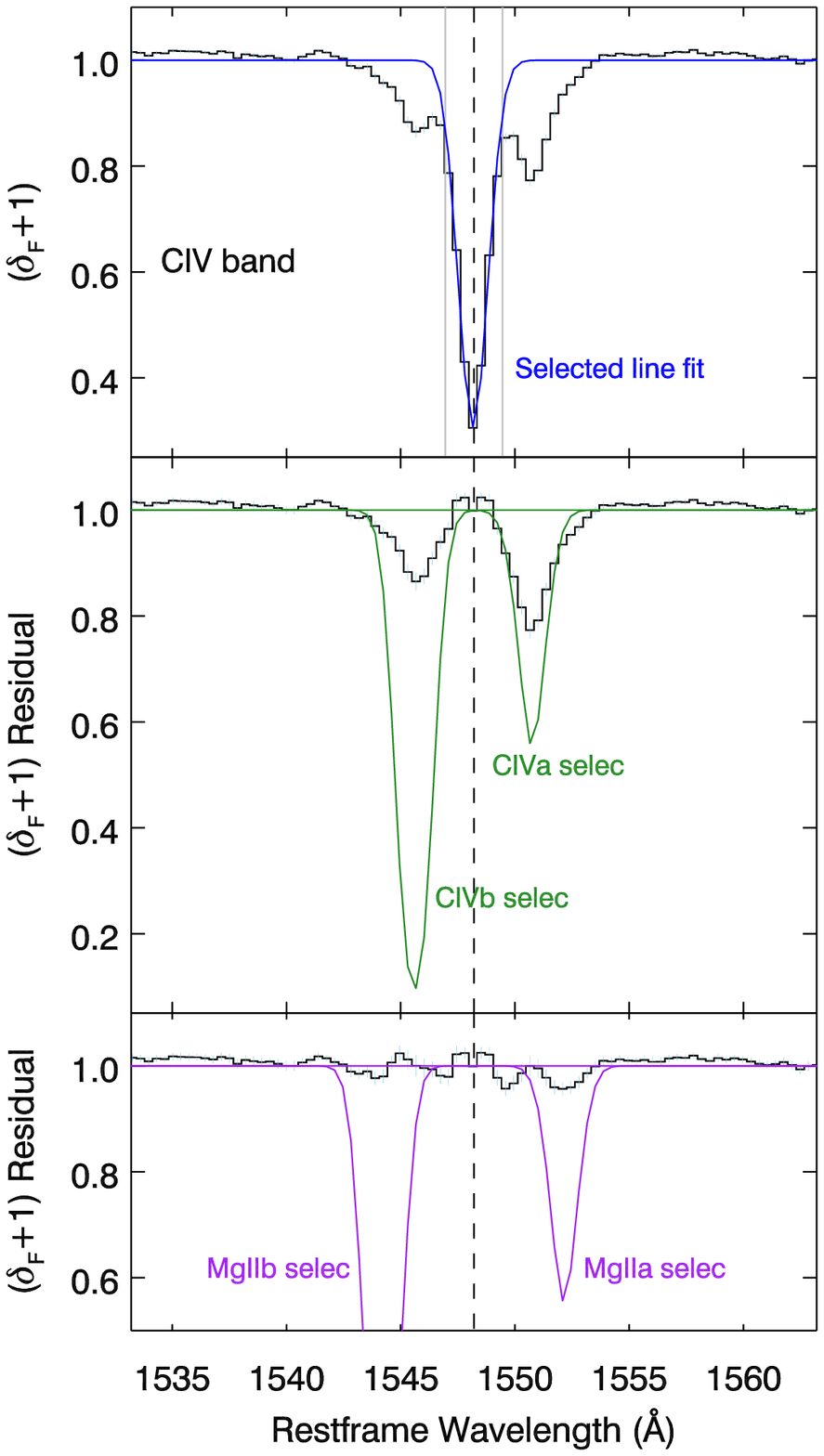}
\includegraphics[angle=0, width=.46\textwidth]{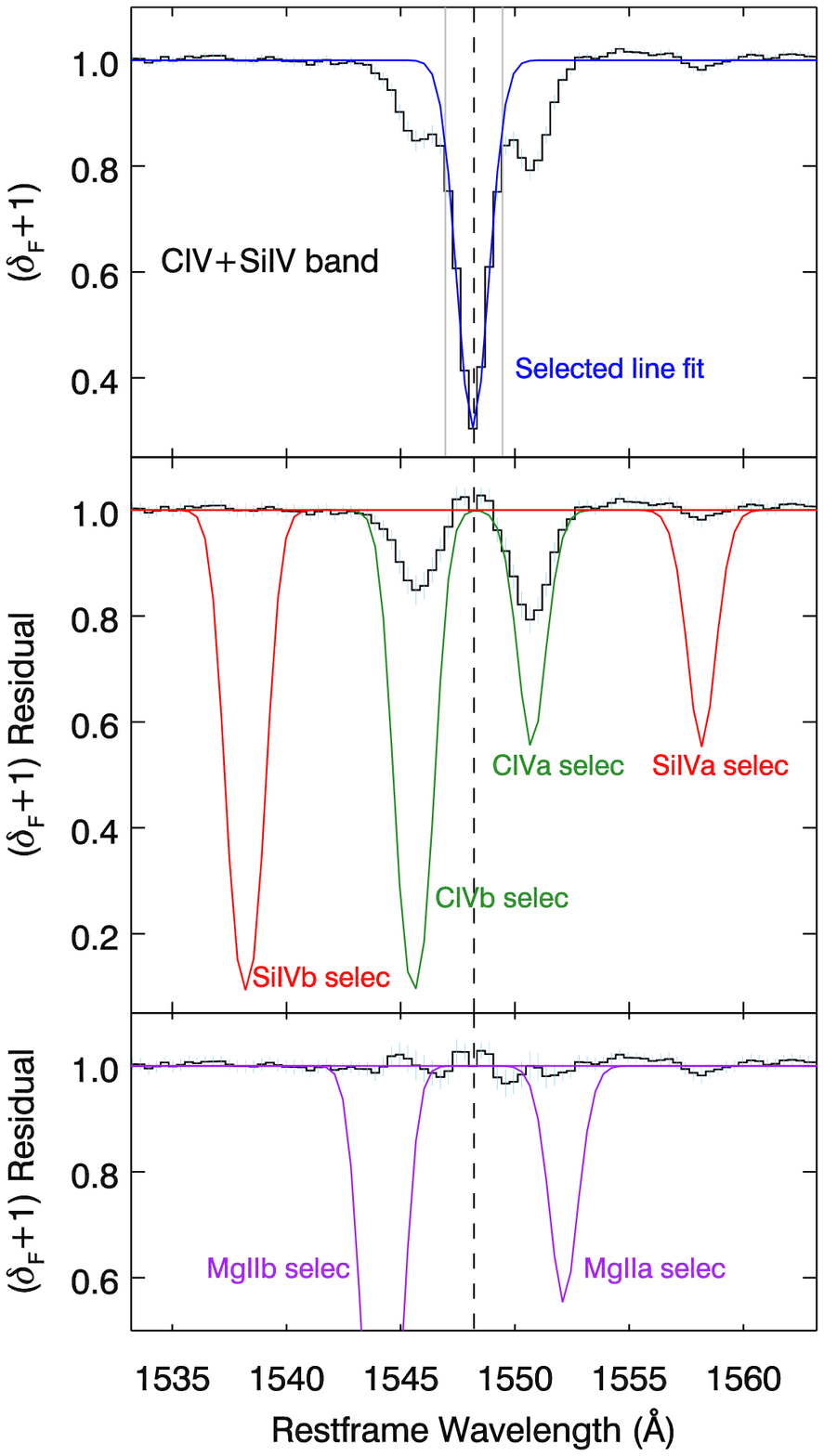}
\end{center}
\caption{Example composite spectra produced by stacking metal absorption with $0.25 \le F < 0.35$ blindly on the red side of quasar \Lya\ emission lines.  The left-hand panels show stacking in the \CIV\ band and the right-hand panels show stacking of absorption in the \CIV+\SiIV\ band. The top panels show the fit (in blue) to the stacked feature (shifted the rest frame wavelength of \CIVa\ and marked with a dashed line). This fit was performed over a region bounded by the grey vertical lines. In the {it middle} panels, the absorption due to the selected feature is corrected for using this fit. Projected absorption based on this fit and arising from 100\% selection of absorbing transitions  are shown as marked. In the {it bottom} panels, signal due to the selection of \CIV\ doublet absorbers is corrected for by rescaling the above projection to match the apparent population sizes, and projected absorption due to the selection of \MgII\ doublet absorption is shown. In each case the difference between projected and observed absorption provides a measure of the fractional incidence of stacking absorbers from the labelled transition.}
\label{blindstack}
\end{figure*}

A key requirement for a viable probe of BAO is the degree of contaminating absorption arising due to other metal absorbers. Unwanted absorption may add both systematic and stochastic noise to a 3D large scale structure measure (\citealt{2011JCAP...09..001S, 2014arXiv1404.1801D}; Bautista et al. in preparation). 

In order to explore the populations of metal absorbers in the two red side spectroscopic bands, I stack pixels with flux  $(0.1n-0.05) \le F < (0.1n+0.05)$ (where $n=0,1,\ldots,9$) amongst the high S/N subset of DR9 including low redshift quasars. I do this blindly with no attempt to limit the source absorption transition. In each case the absorber is assumed to have a rest-frame wavelength of 1548\AA\ such that the observed wavelength is $\lambda_o = 1548(1+z)$ (so the wavelength is scaled by a factor of $1/(1+z)$ before stacking). In the resultant composite spectra, 
the size of the absorbing populations from species such as  \CIV, \SiIV\ and \MgII\ can be ascertained using their doublet lines.
The line strength ratio arising from the exclusive selection of any of these doublet lines is deterministic, and so deviations in the strength of these features in the composite spectrum provide a direct measurement of dilution from other absorbers or noise. For example if \CIVa\ is selected half the time, then one would expect the flux decrement ($D\equiv 1-F$) associated with \CIVb\ to be half that for perfect \CIVa\ selection (with a small correction for the mean opacity of the metal forest)

Fig.~\ref{blindstack} shows an example composite spectra of pixels with a flux between $0.25 \le F < 0.35$.  Model profiles are plotted projecting line strengths assuming 100\% selection of the line labelled. Hence, for a selected transition $\lambda_{\rm selec}$ and its correlated transition $\lambda_{\rm correl}$ the resultant feature would be present at 
 $\lambda_{\rm correl} \lambda_{\rm CIVa} /\lambda_{\rm selec}$. The top panels of Fig.~\ref{blindstack} shows the flux contrast in the \CIV\ band (left) and the \CIV+\SiIV\ band (right). In the regions bounded by the grey vertical lines, a model Gaussian flux decrement profile was fitted to the feature arising due to absorber selection, shown as a blue line.
The middle panels of Fig.~\ref{blindstack} show the residual $\delta_{F}$  when the fitted absorption profiles have been corrected for in both red side bands (by adding fitted flux decrements to the flux contrast in the top panels). In the \CIV\ band, projected doublet line signal associated with perfect selection of both \CIV\ doublet members are shown. In the \CIV+\SiIV\ band, this is repeated with the addition of  projections for the \SiIV\ doublet again assuming perfect selection. The bottom panels of these figures show the residual $\delta_{F}$ corrected for the \CIV\ doublet absorption shown in the middle panels and
 rescaled to the observed \CIV\ absorber fractions in the composite. The projected absorption due to perfect selection of \MgII\ doublet members is shown here.

Using the above procedure, the fraction of doublet lines from \CIV, \MgII, and \SiIV\ were measured. Contributions from other red side metals such as \AlII, \AlIII\ and \FeII\ were explored, but no indications of their presence in the composite were found. Hence, we shall assume that all selected pixels not identified as one of the above species 
arise due to noise. 

Fig.~\ref{absfrac} shows the measured absorber fractions in our sample in both red side bands derived from this analysis. An uncertainty in the residual $\delta_{F}$ of approximately 0.03 is introduced by the imperfect fitting and removal of absorption and this leads to an uncertainly of approximately 0.05 in the absorber fractions that result. As such measurements of \SiIVb\ and \MgII\ may be viewed as upper limits.  In both red side bands the strongest carbon doublet member \CIVa\ dominates the metal absorber population, representing 60-70\% in the \CIV\ band and 50-70\% in the \CIV+\SiIV\ depending on the absorber strength. In all but the strongest absorbers, the next largest contribution comes from the weaker member of the \CIV\ doublet. When combined the \CIV\ doublet make up around 80\% of absorbers in both bands.

\section{A Metal BAO survey}
\label{survey}

\subsection{\Lya\ and \CIV\ as structure tracers}

\CIVa\ is the dominant absorption species in the \CIV+\SiIV\ and \CIV\ bands. This constitutes a maximum redshift path of $\Delta z=0.5$ at z=2, which is comparable to the redshift path of \Lya\ BAO measurements. The bands described here are somewhat narrow and might be expanded given testing, particularly into the \SiIV\ emission line region defining a gap between bands. 

\CIVb\ as the most significant additional absorbing population may contribute to the BAO measurement signal. The restframe wavelengths differ by 0.2\%, much smaller than the anticipated BAO scale precision, hence the \CIVb\ signal can be used to boost the signal (although the effective rest-frame wavelength of the combined doublet must be determined mocks to ensure that a systematic error is not introduced).
Even a small contribution from \SiIV\ might not prove to be a significant contaminant
 as the line-of-sight BAO scale at a given observed wavelength is approximately equal in \CIV\ and \SiIV\ for concordance cosmology. 
 \MgII\ may also present a contaminant, but for concordance cosmology the transverse BAO separation differs from that of \CIV\ by a factor of $\sim 3$ making it simple to distinguish.
 Alternatively, one might seek to fold these additional absorption transitions into a measurement using the approach set out by \citet{2013JCAP...09..016I}.

Since metals arise in the intergalactic and circumgalactic media via extragalactic outflows, it is reasonable to expect that metals trace richer environments that typical forest absorbers.
Observations indicate that the volume filling factor of metals need not be 100\% {e.g. \citealt{2003ApJ...596..768S,2004MNRAS.347..985P}) and that galaxy proximity in addition to gas density (and so \Lya\ absorption strength) is a factor in setting the \CIV\ opacity \citep{2006ApJ...638...45P, 2014arXiv1403.0942T}. Furthermore, results from P14 indicate that the strongest \Lya\ forest absorbers (with $\delta_{\rm Ly\alpha}<0.58$) at SDSS resolution and in the absence of noise typically represent circumgalactic regions. Hence, the \CIV\ forest is likely to be more biased than the \Lya\ forest, which would increase the relative clustering amplitude in the \CIV\ autocorrelation.

A further potential benefit of using the \CIV\ forest arises due to its relative dynamic range. The \Lya\ forest saturates when structures reach an over density as low as 10. As a weaker absorbing population as a whole, the \CIV\ forest rarely saturates and has the potential to distinguish higher density systems, which may also lead to a more refined probe of biased systems.

\begin{figure}
\begin{center}
\includegraphics[angle=0, width=.48\textwidth]{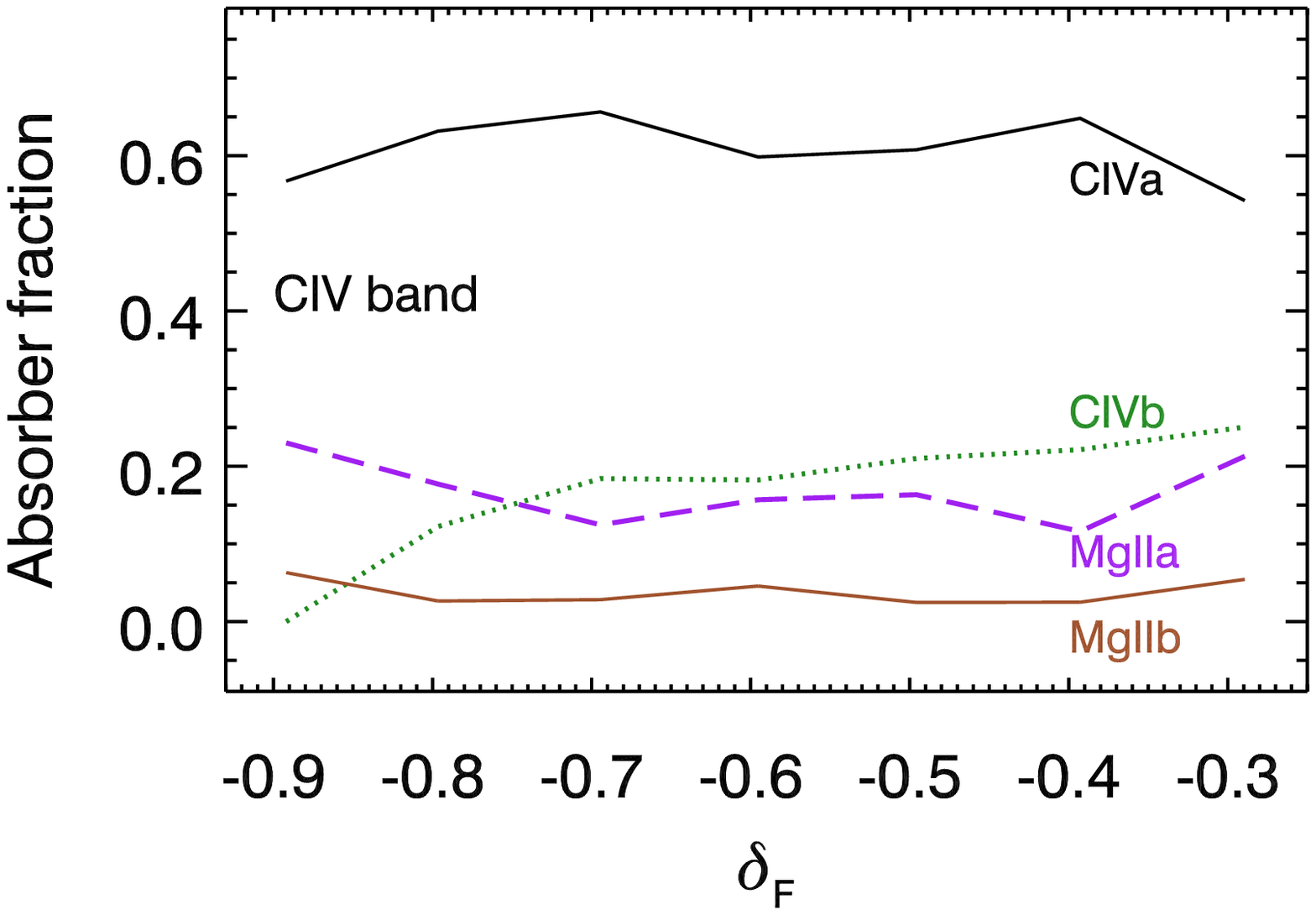}
\includegraphics[angle=0, width=.48\textwidth]{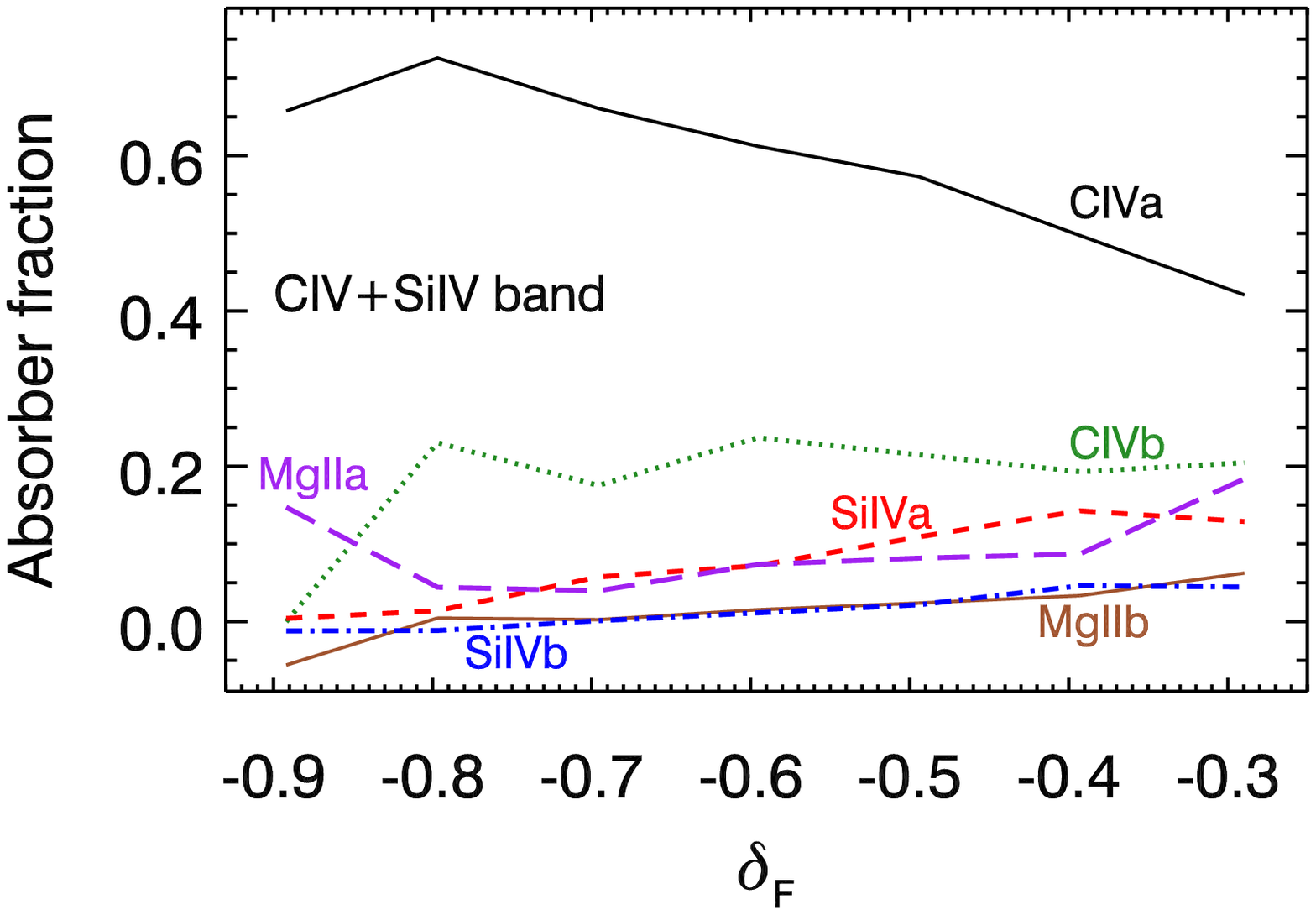}
\end{center}
\caption{The fraction of metal absorbers arising due to the labelled transitions in the \CIV\ band (top) and the \CIV+\SiIV\ band (bottom). Fractions are measured by comparing projected and observed absorption for doublet lines in average composite spectra such as that shown in Figure~\ref{blindstack}. The fraction of  absorbers is presented assuming that all selected pixels without identified associated arise due to stochastic noise (i.e. fractions sum to 1). }
\label{absfrac}
\end{figure}

\subsection{BAO scale precision}

I forego a full Fisher matrix projection of cosmological constraints in light of the uncertainties in the use of the \CIV\ forest as a probe. Instead I use a simple scaling relationship argument. In Section~\ref{strength},
we found that the \CIVa\ forest is a factor of 5--30 weaker than the \Lya\ forest, and since the \CIVb\ forest can be used to boost the signal this becomes a factor of 4--25. This is consistent with the measurement that the total 1D power in the \CIV\ forest is a factor of 5--20 lower that the 1D power in the \Lya\ forest. We may compare these probes directly for a \CIV\ absorption survey with mean redshift $\langle z \rangle \approx 1.6$ compared to \Lya\ forest surveys $\langle z \rangle=2.3$, assuming similar bias, data quality, contaminants, effective volume and path length. Following the formalism of equation 12 of \citet{2011MNRAS.415.2257M} (see also \citealt{2007PhRvD..76f3009M}), a decline in signal power must be accompanied by an increase in quasar number density in order to retain equal precision on the measurement of the BAO scale. 

eBOSS  (with SDSS-II DR7 and BOSS data) is expected to obtain spectra of $1.3<z<2.15$ quasars with approximately four times the number density of BOSS $z> 2.15$ quasars (C. Y{\`e}che, private communication). Quasars with redshifts up to $z=2.7$ will provide useful path for \CIV\ absorption below $z=2$ and since these quasars make up approximately two-thirds of the high redshift BOSS and eBOSS samples, this brings the effective increase in number density to a factor of 5. One might also supplement this probe by extending to the higher redshifts probed by the \Lya\ forest autocorrelation. Therefore, we may conclude that eBOSS will provide a \CIV\ autocorrelation measurement of BAO with precision of order that obtained by the DR11 \Lya\ autocorrelation analysis of \citet{2014arXiv1404.1801D}, assuming all other observing conditions to be equal. This corresponds to a measurement uncertainty of 2\%.

The primary science driver for obtaining this high surface density of low redshift quasars is the measurement of BAO using quasar redshifts. This would probe structure at largely the same epoch as proposed here, but with different systematic errors and observing biases. \citet{2014JCAP...05..027F} demonstrated that much can be gained from cross-correlating quasar redshifts with the \Lya\ forest despite the sparsity of the quasar redshift data in BOSS.

eBOSS will provide a much better sampled quasar survey -- sufficiently high to obtain a 2\% uncertainty of the BAO scale in the eBOSS quasar autocorrelation (www.sdss3.org/future).
As argued by \citet{2014JCAP...05..023F}, in the shot noise dominated limit the fractional error in the BAO scale for a absorber-quasar cross-correlation measurement is the inverse sum of the individual fractional BAO errors from the autocorrelation measurements. Therefore, a \CIV-quasar autocorrelation has the potential to measure the BAO scale with uncertainty as low as 1\%.

\section{Observational challenges}
\label{challenges}

Renormalizing quasar spectra to correct for the unabsorbed continuum level represents a significant challenge to performing cosmology in data of this quality. The use of an absorption probe with signal reduced by an order of magnitude or more might naively be presumed to preclude a useful measurement. However, the high incidence rate of \Lya\ absorption in high redshift studies creates limitations that are not present in the \CIV\ forest. The \Lya\ forest is sufficiently dense by comparison to the BOSS spectral resolution that it is not possible to identify unabsorbed regions with high confidence. Hence alternative approaches are typically used such as joint fits to a stacked mean quasar spectrum and the forest flux probability distribution function \citep{2013A&A...552A..96B} or the production of principal components based on low redshift quasars that lack this dense absorption \citep{2013AJ....145...69L}. 
The \CIV\ forest is sufficiently sparse that identifying unabsorbed regions in the data will be substantially easier, allowing simple spline fits such as those used here. 

Mocks will be required to test a variety of systematics in the data. We may utilize the methods described by Bautista et al (in prep) (see also \citealt{2012JCAP...01..001F,2014arXiv1404.1801D}), but some additionally refinements are necessary in the generation of metal forest mocks. The physics governing the distribution of metals is more complex than the physics governing the distribution of neutral hydrogen. Metals only arise in the intergalactic medium by virtue of extragalactic mechanical outflows and it is an open question as to how they are distributed with respect to faint (typically unobserved) high redshift galaxy populations (e.g. \citealt{2014MNRAS.438..529R}). 

Neutral hydrogen is thought to trace large-scale structure to first order, but observed metal absorption is far from ubiquitous.  As discussed in Section~\ref{survey}, observations indicate that the metal enrichment is non-uniform. Also different metal ionization species are thought to trace different subsets of metal enriched gas. The patchy nature of the \CIV\ forest may pose a challenge in the production of mock spectra. However,  absorption based BAO measurements are thought to be largely insensitive to small scale gas physics \citep{2011MNRAS.415.2257M} and it is possible to explore a range of metal population scenarios in mocks to test BAO measurement and bias is affected.

An effect that must be explored are fluctuations in the extragalactic ionizing UV background. The different ionization potential of \CIV\ compared to \HI\ and the extra ionization states in carbon complicate the ionization characteristics of \CIV. Further potential effects may be present if the \CIV\ absorption tends to arise in regions close to galaxies with excess UV levels compared to the background. Also inhomogeneity in the \CIV\ fraction may arise on large scales due to post helium reionization proximity to quasars. The severity of this effect is somewhat limited (Pieri et al., in preparation), but it is notable that such an effect may impact upon the feasibility of quasar-absorption cross-correlation using \CIV.

\section{Summary}

I have explored the potential for using the \CIV\ forest down to redshift $z=1.3$ as a new probe of baryon acoustic oscillations, and an alternative to the \Lya\ forest. The need for turning to this weaker probe of intergalactic structure is driven by the fact that \lya\ absorption drops out of the optical window at $z\approx 2$, and the relative ease with which quasars below $z\approx2$ can be surveyed.

Two bands redward of the quasar \Lya\ emission line were investigated between \Lya\ and \CIV\ in emission, split by the \SiIV\ emission line. Using BOSS DR9 quasar spectra, the metal power, metal populations, and strength of the \CIV\ absorption signal in relation to \Lya\ absorption were tested. I find that \CIV\ absorption dominates these bands, and represents 80\% of absorption. Furthermore, the signal associated with the \CIV\ forest at low redshift is a factor of 4--25 weaker that that associated with the \Lya\ forest at high redshift, but with greater bias, and a dynamic range suited to probing higher density regions.

The upcoming eBOSS survey will provide five times the effective number density of quasars available for a $z<2$ \CIV\ forest survey compared to the BOSS \Lya\ forest survey. Therefore it may provide broadly comparable accuracy of around 2\% in the BAO scale. In combination with the eBOSS quasars redshift distribution, the \CIV- quasar cross-correlation may provide uncertainty as low as 1\%. Forthcoming quasar surveys as part of Dark Energy Spectroscopic Instrument (DESI) and WHT Enhanced Area Velocity Explorer (WEAVE) also have the potential to provide improved constraints.

\section*{Acknowledgments}

I thank Andreu Font-Ribera, Pat McDonald, Will Percival and Stephan Frank for valuable conversations, Jim Rich for his comments on the manuscript, and Christophe Y\`eche for providing helpful information about eBOSS.

The research leading to these results has received funding from the European Union Seventh Framework Programme (FP7/2007-2013) under grant agreement no. [PIIF-GA-2011-301665]. 

Funding for SDSS-III has been provided by the Alfred P. Sloan Foundation, the Participating Institutions, the National Science Foundation, and the US Department of Energy Office of Science. The SDSS-III web site is http://www.sdss3.org/.

\label{lastpage}

\end{document}